\documentclass[preprint,amsmath,amssymb,apl]{revtex4}

\usepackage{graphicx}
\usepackage{dcolumn}
\usepackage{bm}
\usepackage{amssymb}

\begin{document}

\title{Inverted spin polarization of Heusler alloys for new spintronic devices}
\author{Andy Thomas}
\email{athomas@physik.uni-bielefeld.de}

\author{Dirk Meyners} %
\author{Daniel Ebke} %
\author{Ning-Ning Liu} %
\author{Marc D. Sacher} %
\author{Jan Schmalhorst} %
\author{G\"unter Reiss} %
\affiliation{Bielefeld University, Thin Films \& Nanostructures, Bielefeld, Germany}
\author{Hubert Ebert}
\affiliation{Department of Chemistry and Physical Chemistry, Ludwig-Maximilians-University of Munich, Germany}
\author{Andreas H\"utten}
\email{Andreas.Huetten@fzk.de}
\affiliation{Bielefeld University, Thin Films \& Nanostructures, Bielefeld, Germany}
\affiliation{Research center Karlsruhe GmbH, Institute for Nano-technology, 76021 Karlsruhe, Germany}


\date{\today}


%
\begin{abstract}
A new magnetic logic overcomes the major limitations of field programmable gate arrays while having a 50\% smaller unit cell than conventional designs utilizing magnetic tunnel junctions with one Heusler alloy electrode. These show positive and negative TMR values at different bias voltages at room temperature which generally adds an additional degree of freedom to all spintronic devices. 
\end{abstract}
\maketitle
The logic circuits of computers are based on integrated transistors (ICs)\cite{ic_kilby}. These circuits can generally not be altered during run-time, but are designed as general purpose central processing units (CPUs)\cite{neumann}. Because of this limitation, specialised ICs are introduced for tasks such as graphic or sound processing. Although alterable ICs exist, the so-called field programmable gate arrays (FPGAs) have slow switching and operational speeds. Therefore, they are not reprogrammed during run-time\cite{apl2002V80S1291,ieee2002V90S1201}. Magnetic tunnel junctions (MTJs) exhibit the tunnel magneto resistance (TMR) effect\cite{PL1975V54AS225} and can be used as a base for magnetic memory or logic\cite{sci1998V282S1660} to beat these shortcomings.

An MTJ consists of two metallic, ferromagnetic layers serving as electrodes separated by a thin insulating layer acting as a tunnel barrier\cite{MMM1999V200S248}. The resistance of this device is generally high when the two magnetization directions are anti-parallel. This state ($\rm R_{ap}$) can be identified as logic 1 and the parallel alignment ($R_p$) as 0. To have a large separation between the two logic states a worldwide race after large tunnelling magnetoresistance (TMR) values (which is the difference of the two resistances ($\rm R_{ap}-R_p)/R_p$) was initiated. 

The current successor in this competition is a combination of FeCoB electrodes and MgO tunnel barrier\cite{nmat2004V3S868,nmat2004V3S862}. This type of MTJ is capable of reaching TMR-effect amplitudes of larger than 200\% at room temperature (RT). In order to increase this TMR value further, half-metallic ferromagnets with a band gap at the Fermi level ($E_F$) -- resulting in 100\% spin polarization and possible TMR values of several thousand percent -- are being tested as new electrode materials. To date four materials classes with 100\,\% spin polarization have theoretically been predicted: Oxide compounds\cite{jap2002V91S8345}, perovskites\cite{MMM1997V172S237}, zinc-blende compounds\cite{JJAP2000V39S1118} and Heusler alloys\cite{prl1983V50S2024}. In particular full-Heusler $\rm X_2YZ$ alloys (X, Y: transition metal elements, Z: group III, IV, or V element) have a high potential to realise half metallic behaviour at room temperature (RT) due to high Curie temperatures. 

Recently, very high TMR effect amplitudes of $\rm Co_2MnSi/AlO_x/Co_{70}Fe_{30}$ at RT were already shown\cite{APL2005V86S052501}. But the large TMR values are not the main advantage of using Heusler MTJs as base structures in logic circuits or other spintronic devices. The unique band structure of $\rm Co_2MnSi$/$\rm Co_2FeSi$ enables us to boost the progress in programmable logic based on MTJs. 

The magnetic tunnel junctions were prepared using dc and rf magnetron sputtering in an ultra high vacuum chamber. The full layer stack was $\rm Si(001)/ SiO_x 50\,nm/ V 42\,nm/ Co_2MnSi 100\,nm/ Al 2.3\,nm + oxidation/ Co_{70}Fe_{30} 5\,nm/ Mn_{83}Ir_{17} 10\,nm/ Cu 40\,nm/ Ta 5\,nm/ Au 40\,nm$. The stack was in-situ annealed after Aluminium oxidation (40\,min/ 400\,C - 500\,C) to induce texture and atomic ordering of the $\rm Co_2MnSi$ layer. The whole layer stack was ex-situ annealed for 1\,h at 275\,C in an external magnetic field of 100\,mT to establish the exchange bias between Ir-Mn and Co-Fe. It was patterned by optical lithography and ion beam etching yielding quadratic junctions between $\rm 300 \times 300\mu m^2$ and $\rm 5 \times 5\mu m^2$. The transport properties were determined using conventional 2 point dc technique, current crowding effects were carefully excluded\cite{APL1967V10S29}. The band structures were calculated utilising the SPR-KKR program package\cite{sprkkr}.

\begin{figure}
\includegraphics[width=80mm]{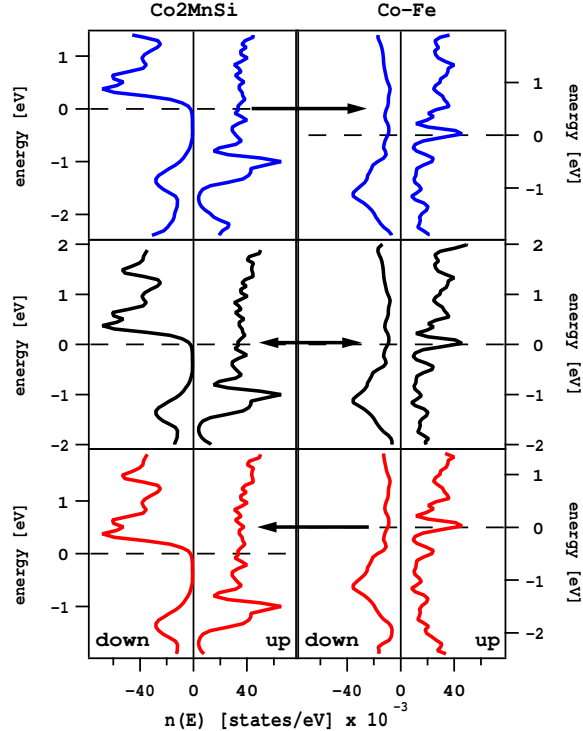}
\caption{\label{fig:band}Calculated band structure of s-electrons of $\rm Co_2MnSi$ and $\rm Co_{70}Fe_{30}$ \cite{sprkkr}. Please note the gap in the $\rm Co_2MnSi$ spin down density of states and, subsequently, the sudden increase at positive energies. This leads to positive spin polarisation at zero bias and negative values above 1200\,mV in the experiment. The blue, black and red curves show a bias voltage of +500\,mV, zero, -500\,mV, respectively.}
\end{figure}
The key feature of the density of states (DOS) of the tunnelling s-electrons\cite{PRB1973V8S3252} (the tunnel barrier material\cite{SCI1999V286S507} and structure\cite{jap2005V97S10C908} can select different electrons) of $\rm Co_2MnSi$ is shown in figure \ref{fig:band}. For the spin down electrons there is not only a gap just below the Fermi energy but a pronounced peak located just above the band gap. We apply the definition of the spin polarisation
$
P=\frac{N(E)_\uparrow-N(E)_\downarrow}{N(E)_\uparrow+N(E)_\downarrow}
$
where  $N(E)_\uparrow$ and $N(E)_\downarrow$  are the DOS of majority and minority electrons, respectively. It becomes clear that the resulting spin polarization of $\rm Co_2MnSi$ is negative for energies well above 400\,mV.
\begin{figure}
\includegraphics[width=80mm]{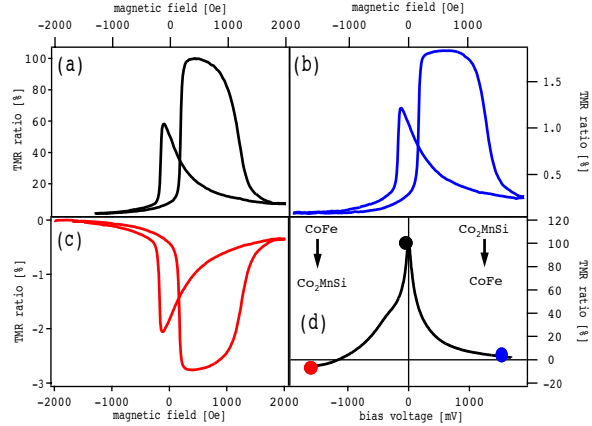}
\caption{\label{fig:loops}(d) Typical bias voltage dependence of the TMR ratio: (a) Positive TMR at zero bias. (b) Positive TMR at +1900\,mV. (c) Negative TMR at -1300\,mV. }
\end{figure}
Figure \ref{fig:loops}d does not become negative at 400\,mV because the integral up to a particular energy determines the effective spin polarisation. Nevertheless, the characteristic kink in the bias voltage dependence is clearly seen. In contrast, the spin polarization of the $\rm Co_{70}Fe_{30}$ counter electrode is positive within the same energy range due to the dominating majority spin population.

Applying a negative bias voltage to a $\rm Co_2MnSi/AlO_x/Co_{70}Fe_{30}$ MTJ enables us to drive s-like electrons from occupied states at $E_F$ of Co-Fe through the Alumina tunnel barrier into unoccupied s-like states of the $\rm Co_2MnSi$ above its Fermi-level. The TMR effect is generally interpreted by Julli\`ere`s expression:
$
\frac{\Delta R}{R} = \frac{R_{AP}-R_P}{R_P}=\frac{2P_1P_2}{1-P_1P_2}
$
relating $R_{AP}$ and $R_P$ to their associated spin polarisations $P_1$ and $P_2$, $P_{\rm Co-Fe}$ and $P_{\rm Co_2MnSi}$ in our case. It becomes obvious that the resulting TMR-effect should be negative for high negative bias voltages (below -1.2\,V in the experiment). This can clearly be seen in the bias voltage dependence of the TMR ratio above RT (at 20\,K given in fig. \ref{fig:loops}). For positive bias voltage the dominating population of the spin up states of both electrodes leads to a positive TMR effect amplitude which decreases with increasing bias voltage\cite{MMM1999V200S248}. A large TMR effect amplitude is expected at zero bias voltage due to the energy gap in the minority DOS of $\rm Co_2MnSi$. This is also experimentally observed reaching more than 100\,\% TMR ratio (fig. \ref{fig:loops}). Consequently, we can select negative and positive magnetoresistance by applying different bias voltages.

The use of these magnetoresistive elements in logic circuits is very promising, due to the non-volatile information storage even without any supply of power. A number of concepts was proposed to build field programmable gate arrays (FPGA) with MTJs. Ney\cite{nat2003V425S485} proposed a logic where only one MTJ could be switched between AND, OR, NAND and NOR. This dramatically decreases the size of a 'unit cell' on the silicon wafer die, but the approach needs a stable, but switchable hard electrode  which is very difficult to achieve and an additional processor cycle to initialise the gates. Hassoun\cite{ieee1997V33S3307} proposed another concept and omitted the second cycle, but the definition of the output signal had to change from 0 to 1 (which is always possible). Furthermore, 4 MTJs were necessary for one gate. 

\begin{figure}
\includegraphics[width=80mm]{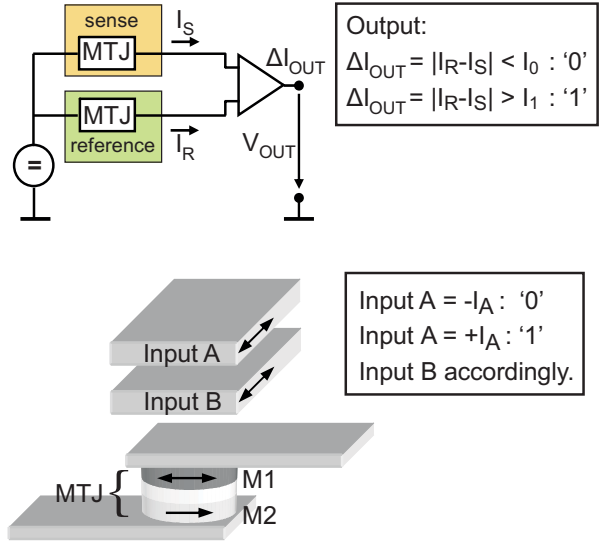}
\caption{\label{fig:circuit}(top) Electrical circuit of the two magnetic tunnel junctions without input leads and definition of the output function. (bottom) Programming of the junctions and input definition. M2 is the fixed and M1 the free electrode.}
\end{figure}
\begin{figure}
\includegraphics[width=80mm]{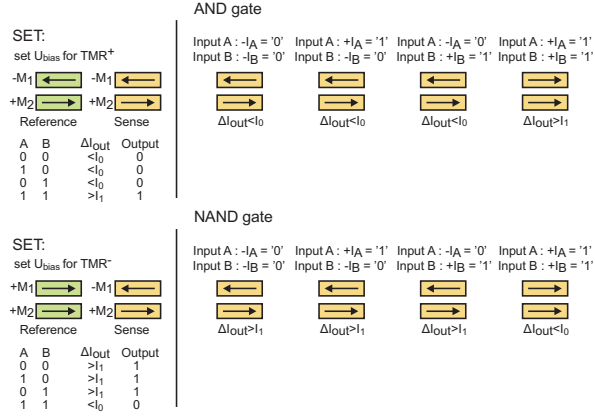}
\caption{\label{fig:AND}Set the magnetic alignment and the applied bias voltage to define the logic function: (top) AND (bottom) NAND; analogous selection of all other functions. The magnetisation of the free electrode can only switch if both input currents have the same sign.}
\end{figure}
Our proposal uses 2 MR-elements, needs no programmable hard electrode, but can be switched between all 4 mentioned functions. This is achieved by using the unique property of negative TMR at room temperature of the presented Heusler alloy. 
Figure \ref{fig:circuit} shows the scheme of the FPGA array based on magnetic tunnel junctions. The sense and reference MTJ are operated using a constant voltage source. The two currents $I_S$ and $I_R$ are compared, this gives the logical output of the gate. The logical input and the programming of the desired function are done by short pulses through the leads that align the soft magnetic electrodes of the MTJs. The actual switch of the logic function from AND to NAND is shown in figure \ref{fig:AND}.
The other functions are selected in the same way by switching the reference cell and the applied bias voltage.

\begin{figure}
\includegraphics[width=80mm]{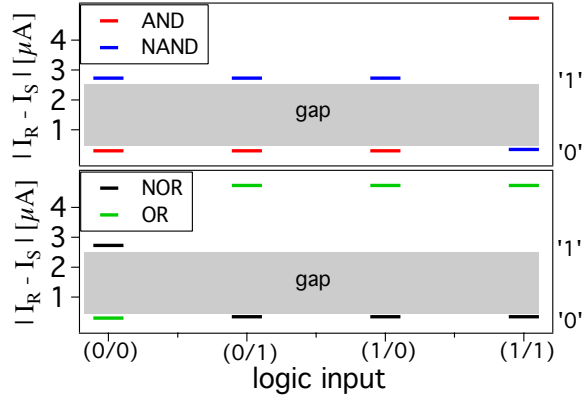}
\caption{\label{fig:functions}Demonstration gate of two $\rm Co_2FeSi/AlO_x/Co_{70}Fe_{30}$ magnetic tunnel junctions switched to the desired logic function and measured in the remanent state at room temperature. The $\rm Co_2FeSi$-Heusler alloy was used due to the minority peak closer to the Fermi-energy and therefore a lower operating voltage of $\pm1034$\,mV. }
\end{figure}
Figure \ref{fig:functions} shows an actual implementation of a Heusler alloy logic gate based on only 2 MTJs. This greatly reduces cost by saving space on the wafer die while maintaining all advantages of other concepts such as non-volatility and very high switching frequencies in the GHz range\cite{nat2002V418S509}. The high frequencies allow a reconfiguration even at run-time and since magnetic random access memory cells (MRAM) are very similar\cite{sci1998V282S1660}, it is even possible to use certain parts of the circuits as logic and others as memory. Furthermore, the negative TMR effect is not limited to $\rm Co_2MnSi$, but was also experimentally observed at room temperature in $\rm Co_2FeSi$ which is used in figure \ref{fig:functions}. Band structure calculations show the sharp increase of the DOS of minority s-electrons just above the band gap in $\rm Co_2FeSi$ and $\rm Co_2FeAl$ as well\cite{Andreas_Heusler}. Stress tests of the logic devices lead to an average life time of the used MTJs of more than 40 years. This is significantly larger than the life span needed in semiconductor industry of more than 10 years\cite{APL2002V80S2335}. 

Finally, the presented logic is only one concept to take advantage of the addition degree of freedom of inverted TMR. This is technologically interesting due to the operation of the metallic system at room temperature (cp. \cite{sci1998V282S1660}). We hope to give input for more devices as well as basic investigations of the unique electronic structure of full Heusler alloys.
\begin{acknowledgements}
This work was supported by DPG funding.
\end{acknowledgements}
%

%
%
%
\end{document}